\documentclass[11pt,a4paper]{article}

\newlength{\colsize}
\newlength{\colsizell}
\usepackage{overcite}
\usepackage{graphicx}
\usepackage{amsmath}
\usepackage{amssymb}
\usepackage{bm}

\newcommand{\la}{\left<}
\newcommand{\ra}{\right>}
\newcommand{\be}{\mbox{$b$}}
\newcommand{\Pone}{\mbox{$P$}}
\newcommand{\Poneasym}{\mbox{$P_{\rm{a}}$}}
\newcommand{\Ponecyc}{\mbox{$P_0$}}
\newcommand{\Ponedil}{\mbox{$P_0$}}

\newcommand{\cP}{\mbox{$c_{\rm P}$}}
\newcommand{\cinf}{\mbox{$c_{\infty}$}}
\newcommand{\veff}{\mbox{$\tilde{v}$}}
\newcommand{\rstar}{\mbox{$r^{*}$}}
\newcommand{\kB}{\mbox{$k_{\rm B}$}}
\newcommand{\bdil}{\ensuremath{b_0}}
\newcommand{\nudil}{\ensuremath{\nu_0}}
\newcommand{\omegadil}{\ensuremath{\omega_0}}
\newcommand{\nmon}{\ensuremath{n_\mathrm{mon}}}

\newcommand{\Icyc}{\ensuremath{I_0}}
\newcommand{\Ione}{\ensuremath{I_1}}
\newcommand{\Itwo}{\ensuremath{I_2}}
\newcommand{\hone}{\ensuremath{h_1}}
\newcommand{\htwo}{\ensuremath{h_2}}
\newcommand{\ctwo}{\ensuremath{c_2}}

\newcommand{\kcut}{\ensuremath{k_c}}

\newcommand{\Acal}{\mbox{$\cal A$}}
\newcommand{\Fcal}{\mbox{$\cal F$}}
\newcommand{\Lcal}{\mbox{$\cal L$}}

\newcommand{\Nav}{\ensuremath{\la N \ra}}

\newcommand{\qvec}{\bm{q}}
\newcommand{\qnulvec}{\bm{q}_0}
\newcommand{\qonevec}{\bm{q}_1}

\newcommand{\qone}{q_1}
\newcommand{\kvec}{\bm{k}}
\newcommand{\rvec}{\bm{r}}
\newcommand{\lvec}{\bm{l}}
\newcommand{\lnulvec}{\bm{l}_0}
\newcommand{\lonevec}{\bm{l}_1}

\newcommand{\lhat}{\ensuremath{\bm{e}}}
\newcommand{\lnulhat}{\ensuremath{\bm{e}_0}}
\newcommand{\lonehat}{\ensuremath{\bm{e}_1}}

\newcommand{\lKuhn}{\ensuremath{l_{\rm K}}}
\newcommand{\sKuhn}{\ensuremath{s_{\rm K}}}
\newcommand{\rhoKuhn}{\ensuremath{\rho_{\rm K}}}

\begin{document}

\title{Distance dependence of angular correlations in dense polymer solutions}

\author{\bf \normalsize
  J.P.~Wittmer,$^\dagger$\thanks{Corresponding author: {\tt joachim.wittmer@ics-cnrs.unistra.fr}}
  A.~Johner,$^\dagger$
  S.~P.~Obukhov,$^\ddagger$ 
  H.~Meyer,$^\dagger$
  A.~Cavallo,$^\dagger$ and
  J.~Baschnagel$^\dagger$
}
  
\date{{\em \normalsize
  $^\dagger$Institut Charles Sadron, Universit\'e Strasbourg, CNRS UPR 22,\\ 
  23 rue du Loess--BP 84047, 67034 Strasbourg Cedex 2, France \\
  $^\ddagger$
  Department of Physics, University of Florida, Gainesville FL 32611, USA} \\[3mm]
\begin{minipage}[t]{0.9\textwidth}
{\small
Angular correlations in dense solutions and melts of flexible polymer chains are 
investigated with respect to the distance $r$ between the bonds by comparing quantitative 
predictions of perturbation calculations with numerical data obtained by Monte Carlo 
simulation of the bond-fluctuation model.
We consider both monodisperse systems and grand-canonical (Flory-distributed) 
equilibrium polymers. Density effects are discussed as well
as finite chain length corrections.
The intrachain bond-bond correlation function $\Pone(r)$ is shown to decay as 
$\Pone(r) \sim 1/r^3$ for $\xi \ll r \ll \rstar$ with
$\xi$ being the screening length of the density fluctuations
and  $\rstar \sim N^{1/3}$ a novel length scale increasing slowly
with (mean) chain length $N$.
%
%
%
%
}
\end{minipage}
}

\maketitle

\section{Introduction}
\label{sec:intro}
%
\begin{figure}
\begin{center}
\centerline{\includegraphics*[width=0.95\colsize]{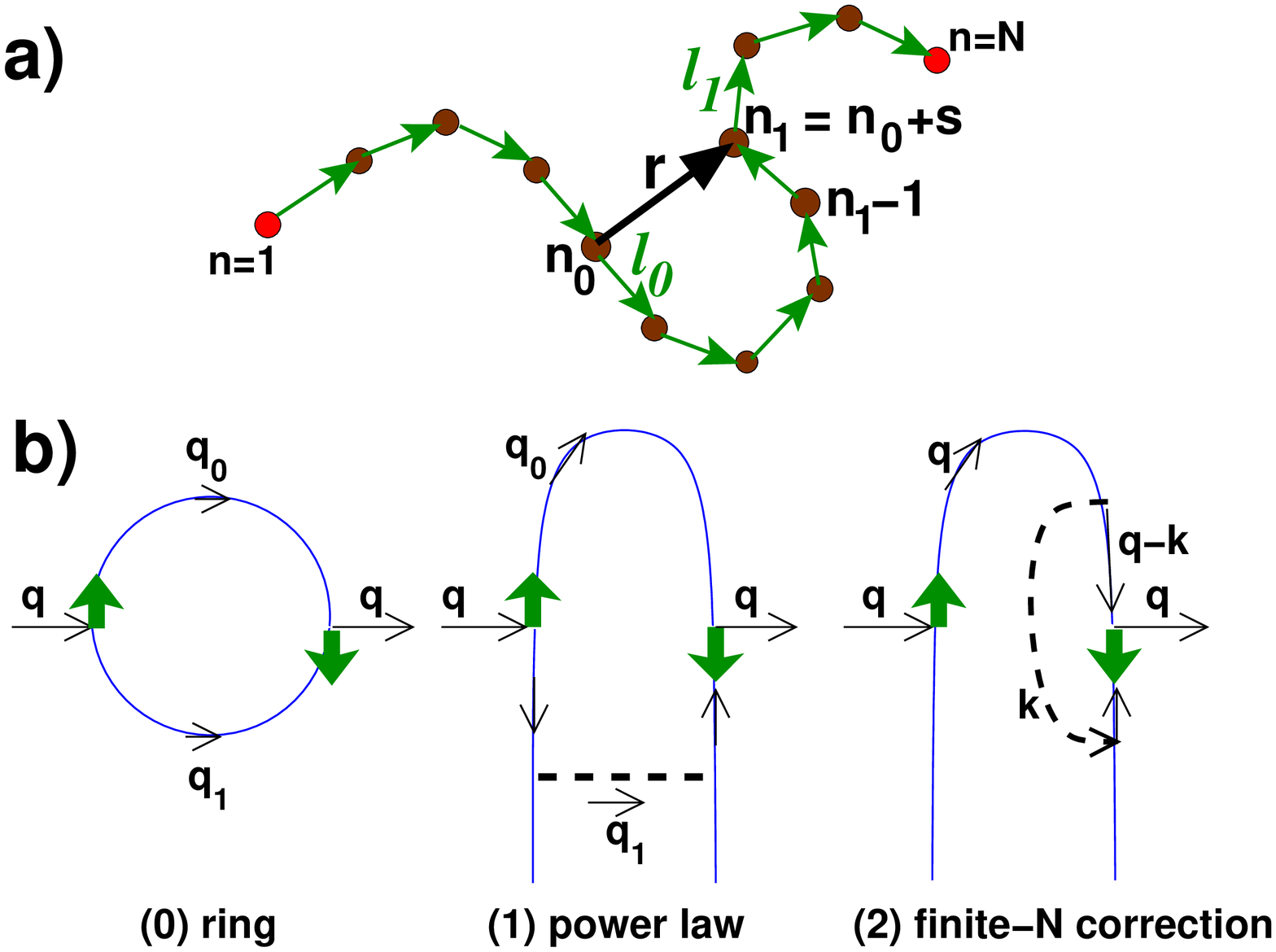}}
\caption{
Sketch of the considered problem in real space {(a)} and of
the first-order perturbation interaction diagrams in reciprocal space {(b)}.
{(a)}
The angular correlations are characterized by the bond-bond correlation function
$\Pone(r) = \la  \lnulvec \cdot \lonevec \ra/l^2$
averaged over all pairs of bonds $\lnulvec$ and $\lonevec$ of a chain of same distance
$r = |\rvec|$ and normalized by the mean-square bond length $l^2$.
{(b)}
The bold vertical arrows represent the bond vectors (eq~\ref{eq_boost}),
dashed lines the effective monomer interactions $\veff(k)$ and
thin lines the Fourier-Laplace transformed Gaussian propagators $G(k,t)$
with the Laplace variable $t$ being conjugated to the curvilinear distance $s$.
The inserted wavevector $\qvec$ is conjugated to the distance $\rvec$
between both bonds in real space.
Angular correlations in ring polymers are described by diagram $(0)$.
The asymptotic behavior of long linear chains (eq~\ref{eq_P1rkey}) 
is obtained from diagram $(1)$.
The last diagram $(2)$ describes the finite-size corrections relevant for large
distances $r \gg \rstar(N) \sim N^{1/3}$.
}
\label{fig_graph}
\end{center}
\end{figure}

\paragraph{Background.}
%
It is generally assumed that large scale correlations are screened in dense solutions 
of flexible polymers beyond the local correlation length $\xi$ characterizing the decay 
of the density fluctuations 
\cite{Flory49,DoiEdwardsBook,DegennesBook}.
%
One consequence of this screening hypothesis is that orientational correlations between 
two bonds $\lnulvec$ and $\lonevec$ on the same chain should vanish rapidly for distances
$r\gg\xi$ and for corresponding curvilinear distances $s = n_1 - n_0 \gg g$
with $g$ being the number of monomers spanning the correlation length.
See Figure~\ref{fig_graph} for a sketch of the notations used in this paper
with $n$ denoting the monomer index,
$N$ the number of monomers per chain,
$R(N)$ the root-mean-square end-to-end distance, 
$\rho$ the monomer number density,
$l$ the root-mean-square bond length,
$\be \equiv \lim_{N\to\infty} R(N)/N^{1/2}$ the effective bond length
and $\cinf = (\be/l)^2$ the dimensionless chain stiffness parameter
\cite{DoiEdwardsBook,foot_asympN}.

Surprisingly, recent numerical studies \cite{WMBJOMMS04,WBCHCKM07,WBM07,papRouse,WCK09}
have demonstrated the power-law decay of the intrachain bond-bond correlation function 
$\Pone(s) \equiv \la \lnulvec \cdot \lonevec \ra/l^2$,
averaged over bond pairs of same curvilinear distance $s$,
as a function of $s$
\begin{equation}
\Pone(s) \approx 
\cP
s^{-\omega}
\mbox{ for }  g \ll s \ll N
\label{eq_Psomega}
\end{equation}
with an exponent $\omega=3/2$ rather than the exponential cut-off expected
from Flory's ideality hypothesis \cite{foot_Ponedef}.
(The amplitude $\cP$ is given in eq~\ref{eq_cP} below.)
%
This result has been rationalized by means of scaling arguments and 
perturbation calculations which demonstrate the systematic swelling of the chain
segments 
\cite{ANS03,WMBJOMMS04,WBJSOMB07,BJSOBW07,WBM07}.
The gist of the calculation is that the effective interactions between the monomers
of a chain are only partially screened and represented (to leading order)
by an effective potential in momentum space
\begin{equation}
\veff(k) \approx \frac{(b k)^2}{12\rho} 
\mbox{ for } \xi \ll 1/k \ll R(N) 
\label{eq_veffasym}
\end{equation}
increasing quadratically with wavevector $k$ \cite{DoiEdwardsBook,BJSOBW07}.
%
The detailed calculation yields the power-law amplitude
\begin{equation}
\cP = 
\sqrt{\frac{3}{8\pi^3}} \frac{\cinf}{\rho \be^3}
\label{eq_cP}
\end{equation}
which is very close to the empirical values for all simulation models tested \cite{WBM07}.

\paragraph{Aim and key results of this study.}
%
Since the power-law decay of $\Pone(s)$ resembles the return probability of
a random walk in three dimensions it is tempting \cite{Rubinstein08} 
to attribute the observed effect to ``self-kicks" of the chain
involving the correlated bonds themselves (or their immediate neighbors). Accordingly,
the bond-bond correlation function should reveal a $\delta(r)$-correlation
if sampled as a function of the distance $r = |\rvec|$
between bond pairs.
This interpretation turns out to be incorrect, however, and we will show that
the power law in $s$ simply translates as
\begin{equation}
s^{-\omega} \Leftrightarrow (r/b)^{-\omega/\nu}
\label{eq_sr_scaling}
\end{equation}
with a Flory exponent $\nu=1/2$ for (to leading order) Gaussian chains.
More specifically, it will be demonstrated by means of analytical theory and 
Monte Carlo simulation that 
\begin{equation}
\Pone(r) 
\approx \Poneasym(r) \equiv \frac{\cinf}{12\pi \rho r^3}
\mbox{ for } \xi \ll r \ll \rstar(N) 
\label{eq_P1rkey}
\end{equation}
as suggested by eq~\ref{eq_sr_scaling},
i.e.~the angular correlations are genuinely long-ranged.
The index ``$\rm{a}$" emphasizes that $\Poneasym(r)$ is the predicted {\em asymptotic} behavior
for long chains on scales where the system
behaves as an incompressible solution. 
For systems with finite compressibility ($\xi \gg b$) the power law
generalizes naturally to 
\begin{equation}
\Pone(r) = \Poneasym(\xi) \ f(u)
\label{eq_P1scaling}
\end{equation}
in terms of the reduced distance $u = r/\xi$ and a scaling function
$f(u) \Rightarrow 1/u^3$ for $1 \ll u$.
%
%
Interestingly, the upper cut-off $\rstar(N)$ of eq~\ref{eq_P1rkey} is found to increase 
rather slowly with chain length 
\begin{equation}
\rstar(N) \approx b N^{1/3} \ll R(N) \approx b N^{1/2}.
\label{eq_rstar}
\end{equation} 
The simulation of computationally challenging chain lengths thus is required
to demonstrate numerically the predicted power-law decay of $\Pone(r)$.

\paragraph{Outline.}
%
The one-loop perturbation calculation leading to the above results
follows again the seminal work by Edwards \cite{DoiEdwardsBook}.
See Figure~\ref{fig_graph}(b) for a sketch of the computed interaction graphs.
This calculation will be discussed first (Section \ref{sec:theo}).
Computational methods and parameters are summarized in section \ref{sec:algo}.
We present then in section \ref{sec:simu} our numerical results obtained for systems 
containing either monodisperse polymers or 
Flory size-distributed equilibrium polymers \cite{WMC98b,foot_annealed}.
By varying the density we scan the screening length $\xi$ over two orders of magnitude \cite{WCK09}. 
This puts us into a position to test the general scaling relation, eq~\ref{eq_P1scaling}, 
for systems with finite compressibility.
Focusing first on the properties of asymptotically long chains we consider 
finally the strong finite chain-size effects predicted by eq~\ref{eq_rstar}.
A synopsis of our results is given in section \ref{sec:summary}
where we suggest possible avenues for future studies.
%

\section{\normalfont \large \bfseries Perturbation calculation}
\label{sec:theo}

\paragraph{General remarks.}
We remind \cite{DoiEdwardsBook} that the first-order perturbation calculation of a 
quantity $\Acal$ under a perturbation $U$ (to be specified below) generally reads 
$\la \Acal \ra \simeq \la \Acal \ra_0 + \la U \ra_0 \la \Acal \ra_0 -\la U \Acal \ra_0$.
%
Averages performed over an unperturbed reference system of Gaussian chains 
of effective bond length $b$ are denoted $\la \ldots \ra_0$. In this study we have to
average the observable $\Acal = \lnulvec \cdot \lonevec / l^2$ 
over all intrachain bond pairs at a given distance $r = |\rvec|$.
For linear chains $\la \Acal \ra_0 = 0$ by construction.
Thus we only have to compute the average 
\begin{equation}
\la \Acal \ra \approx - \la  U \Acal \ra_0
\label{eq_Alin}
\end{equation}
which simplifies considerably the task compared to the perturbation calculation of 
the mean-square segment size $R^2(s)$ 
presented in refs~\citen{WMBJOMMS04,BJSOBW07,WBM07}. 
We remind that for closed cycles the ring closure implies long range 
angular correlations even for Gaussian chains, hence,
\begin{equation}
\la \Acal \ra \approx \la \Acal \ra_0 \ne 0
\label{eq_Acyc}
\end{equation}
to leading order.
We suppose first that the chains are infinite and
on local scale perfectly flexible ($\cinf=(b/l)^2=1$) \cite{foot_bl}. 
We begin by formulating the problem in reciprocal space and
demonstrate then the power-law asymptote, eq~\ref{eq_P1rkey}.
%
%
Finite chain-size effects are first discussed for Flory distributed chains and 
then by means of inverse Laplace transformation for monodisperse melts.
Finally, it is shown how our results can be reformulated for semiflexible chains ($\cinf > 1$).

\paragraph{Reciprocal space description of flexible chains.}
The calculation of eq~\ref{eq_Alin} and \ref{eq_Acyc} is most readily performed in 
reciprocal space as sketched in Figure~\ref{fig_graph}(b).
The Fourier transform of a function $f(\rvec)$ is denoted
$f(\kvec) \equiv \Fcal[f(\rvec)] = \int {\rm d}\rvec f(\rvec) e^{- {\rm i} \kvec\cdot \rvec}$,
the Laplace transform of a function $f(s)$ is written
$f(t) \equiv \Lcal[f(s)] = \int_0^{\infty} f(s) e^{-s t}$
with $t$ being the Laplace variable conjugated to the arc-length $s$.

{\em Bond vectors and observable.}
The vertical bold arrows in Figure~\ref{fig_graph}(b) represent the Fourier transformed bond vectors 
\begin{equation}
\Fcal[\lvec B(\lvec)] =
{\rm i} \bm{\partial_k} B(\kvec) \approx  \frac{{\rm i}}{3} \kvec l^2
\label{eq_boost}
\end{equation}
with $B(\lvec)$ and $B(\kvec) = \Fcal[B(\lvec)]$ being the Gaussian distribution function 
of the bond vector in real and reciprocal space, respectively. We have used here that $B(\kvec)$
can be expanded at low momentum \cite{foot_bondFourier}. The wavevectors conjugated
to the bonds $\lnulvec$ and $\lonevec$ are denoted $\qnulvec$ and $\qonevec$, 
respectively. 
The Fourier transform of the observable $\Acal$ reads, hence, 
\begin{equation}
\Acal(\qnulvec,\qonevec) = - \frac{l^2}{3^2} \qnulvec \cdot \qonevec.
\label{eq_Acal_F}
\end{equation}
%

%
{\em Effective potential $\veff(k)$.}
The perturbation potential $U(\rvec)$ in real space is supposed to be the 
pairwise sum of the effective monomer interactions $\veff(r)$ of all pairs
of monomers of the same chain.
To obtain $\veff(r)$ one labels a few chains.
The interaction between labeled monomers is (partially) screened by the 
background of unlabeled monomers. It has been shown \cite{DoiEdwardsBook,BJSOBW07} 
that within linear response this corresponds to an effective potential 
$\veff(k) = \Fcal[\veff(r)]$ 
with
\begin{equation}
\frac{1}{\veff(k)\rho} = \frac{1}{v\rho} + F(k).
\label{eq_veffgen}
\end{equation}
This potential is represented by the dashed lines in the diagrams.
The bare excluded volume $v$ indicated in the first term of eq~\ref{eq_veffgen}
characterizes the short-range repulsion between monomers. Thermodynamic consistency requires
\cite{ANS05a,BJSOBW07,WCK09} that $v$ is proportional to the inverse of the
measured compressibility of the solution 
\begin{equation}
v =  \frac{1}{g\rho} \equiv \frac{1}{2\rho} \ (a/\xi)^2
\label{eq_vbare}
\end{equation}
where we have introduced a convenient monomeric length $a \equiv b/\sqrt{6}$
and {\em defined} the screening length $\xi$ following, e.g., 
eq~5.38 of ref~\citen{DoiEdwardsBook}. 
Please note that $g$ can be determined experimentally or 
in a computer simulation from the low-wavevector limit of the total 
monomer structure factor and, 
due to this operational definition, $g$ is called ``dimensionless compressibility" \cite{WCK09}.
The single chain form factor $F(k)$ 
represents the interaction between two monomers caused by the 
chain connectivity \cite{DoiEdwardsBook}.
For Gaussian chains of finite length $N$ the form factor is given by Debye's function 
\cite{DoiEdwardsBook}. For infinite chains 
we have, hence, $F(k) = 2/(a k)^2$.
For scales larger than the screening length ($\xi k \ll 1$) the finite
compressibility (indicated by the first term in eq~\ref{eq_veffgen}) becomes
negligible and the solution {\em behaves} for all densities as an incompressible melt. 
Equation~\ref{eq_veffgen} reduces thus to the
scale-free interaction potential already mentioned (eq~\ref{eq_veffasym}).

%
{\em Fourier-Laplace transform of the propagator.}
We remind that the Fourier transform of the Gaussian propagator\cite{DoiEdwardsBook}
may be written $G(k,s) = \exp(-s (a k)^2)$ with $s$ denoting 
the curvilinear distance between two monomers of the chain.
An arrow along a chain contour (thin lines) corresponds to the Fourier-Laplace
transformed Gaussian propagator 
$G(k,t)  = \Lcal[G(k,s)] = 1/((a k)^2 + t)$
of wavevector $k$ as specified in the diagrams.
%
We need to average over all bond pairs at a given distance
irrespective of their curvilinear distance $s$ and we have 
thus to sum below over all possible $s$. For infinite chains this 
corresponds to setting $t=0$ for the corresponding Laplace variable 
and we shall often use the summed up Gaussian propagator
$G(k) \equiv G(k,t=0) = (a k)^{-2}$,
i.e. the Fourier transform of the density $G(r) = 1/(4\pi r a^2)$ 
around a reference monomer of all monomers belonging to the 
same infinite Gaussian chain \cite{DegennesBook}.
Hence,
\begin{equation}
\veff(k) G(k) = \frac{G(k)}{F(k)\rho} = \frac{1}{2\rho}
\label{eq_veff_g}
\end{equation}
for infinite chains on large scales ($\xi k \ll 1$).

{\em Interaction diagrams.}
The momentum $\qvec$ inserted in the interaction diagrams is conjugated to the
distance $\rvec$ between both bonds. Momentum is a conserved quantity flowing
from one correlated point to the other. If the momentum flows in the opposite
direction of a bond (as it is the case for the second bond $\lonevec$)
the wavevector comes with a negative sign in eq~\ref{eq_Acal_F}.
The first two diagrams 
are thus given by the convolution integrals 
\begin{eqnarray}
\Icyc(\qvec) & = & 
   \int_{\qnulvec+\qonevec=\qvec} G(\qnulvec) \Acal(\qnulvec,-\qonevec) G(\qonevec) 
   \label{eq_Icyc} \\
\Ione(\qvec) & = &   
   \int_{\qnulvec+\qonevec=\qvec} G(\qnulvec) \Acal(\qnulvec,-\qonevec) G(\qonevec) 
   \nonumber \\ 
   & & \hspace*{1.3cm} \times [-\veff(\qonevec)] G(\qonevec)
   \label{eq_Ione} 
\end{eqnarray}
The integrals $\Icyc(\qvec)$ and $\Ione(\qvec)$ describe
the bond-bond correlation functions of closed cycles and linear chains,
respectively. 
The perturbation calculation of linear chains using eq~\ref{eq_Alin} 
implies a minus sign. This sign is indicated in front of the effective interaction 
$\veff$ in the last equation.
Using eq~\ref{eq_Acal_F} and assuming eq~\ref{eq_veff_g} 
the integrals are considerably simplified 
\begin{eqnarray}
\Ione(\qvec) & = & - \frac{1}{2 \rho} \ \Icyc(\qvec) 
   \label{eq_Icyc_a} \\
             & = & -\frac{l^2}{18\rho} \int_{\qnulvec+\qonevec=\qvec}  
                    \qnulvec G(\qnulvec) \cdot \qonevec G(\qonevec)
   \label{eq_Ione_a} 
\end{eqnarray}
%
and the inverse Fourier transforms are thus
\begin{eqnarray}
\Ione(r) & = & - \frac{1}{2 \rho} \ \Icyc(r)
   \label{eq_Icyc_b} \\
   & = & \frac{l^2}{18\rho} (\partial_r G(r))^2. 
   \label{eq_Ione_b} 
\end{eqnarray}
\paragraph*{Sum rule for closed cycles and linear chains.}
Up to a constant prefactor the integrals $\Icyc$ and $\Ione$ 
are thus equal on large scales ($r \gg \xi$). 
This remarkable result needs further discussion beyond the technical notions 
set up in the preceding paragraph.

{\em Closed cycles.}
The bond-bond correlation function of closed rings $\Ponecyc(r)$ 
is directly obtained from $\Icyc(r)$
after normalization with the density $G(r)$
\begin{equation}
\Ponecyc(r) = \frac{\Icyc(r)}{G^2(r)} 
= - \left(\frac{l}{3r}\right)^2.
\label{eq_Pcyc_norm}
\end{equation}
The reason for the normalization factor $G^2(r)$ is that 
for $\Ponecyc(r)$ both bonds are {\em known} to be bonds of the same polymer ring 
while the interaction integral eq~\ref{eq_Icyc} corresponds only to a probability 
$G(r)$ for both bonds being in the same chain {\em times}
a probability $G(r)$ that this chain is closed.
That $\Ponecyc(r)$ is negative is of course due to the closure constraint 
which corresponds to an entropic spring force bending the second bond back to
the origin. Since this force is scale free (for infinite chains) this yields a power law. 

{\em Linear chains.}
It follows immediately from eq~\ref{eq_Ione_b} that for linear chains 
\begin{equation}
\Pone(r) = \frac{\Ione(r)}{G(r)} = \frac{1}{12\pi} \frac{1}{\rho r^3}
\label{eq_Plin_norm}
\end{equation}
which demonstrates finally the key claim (eq~\ref{eq_P1rkey}) made in the Introduction
(assuming $\cinf=1$). The normalization factor $G(r)$ is due to the fact 
that for $\Pone(r)$ both bonds are known to belong to the same chain. 
As compared to the closed cycles the correlation has the opposite sign 
since the attractive spring of the ring closure 
(indicated by $G(\qone)$ in eq~\ref{eq_Icyc_a}) 
has been replaced by the effective {\em repulsion} 
(indicated by $-G(\qone) \veff(\qone) G(\qone) = - G(\qone)/2\rho$ in eq~\ref{eq_Ione_a}).
This repulsion bends the second bond away from the origin increasing 
thus the bond-bond correlation function.
{\em Sum rule.}
Interestingly, the perturbation result, eq~\ref{eq_Icyc_b}, may be rewritten as 
\begin{equation}
\Pone(r)  + \frac{G(r)}{2\rho} \Ponecyc(r) =  0
\label{eq_Pr_sumrule}
\end{equation}
where we have used the normalization factors mentioned above.
This ``sum rule" suggests a geometrical interpretation of the observed
relation between infinite linear chains and closed cycles which may 
remain valid beyond the one-loop approximation used here.
The idea is that in an hypothetical ideal melt containing both
linear chains and closed cycles all correlations disappear 
(on distances much smaller than the typical chain sizes)
when summed up over the contributions of both architectures.
The weight $(G(r)/2)/\rho$ corresponds to the fraction of bond pairs
in closed loops \cite{foot_twicecycle}. 
Since the orientational correlations in ideal cycles are necessarily 
long-ranged due the ring closure (eq~\ref{eq_Pcyc_norm}), it follows, 
{\em assuming} the sum rule, that the same applies to bond pairs of linear chains.
Since bonds in closed cycles are anti-correlated ($\Ponecyc(r) < 0$), 
they must be aligned ($\Pone(r)>0$) for linear chains.
\paragraph{Flory size-distributed chains.}
We turn now to the upper boundary indicated in eq~\ref{eq_P1rkey}
and attempt to characterize finite-$N$ effects for $u \gg 1$.
We start by considering self-assembled chains 
(branching of chains and formation of closed loops being disallowed)
having an annealed size-distribution \cite{foot_annealed} with an exponentially
decaying number density $\rho_N = \rho \mu^2 e^{-\mu N}$
for polymer chains of length $N$ with $\mu = 1/\Nav$ being the chemical potential.
This so-called ``Flory size-distribution" is relevant to equilibrium polymer 
systems \cite{WMC98b,WBCHCKM07}.
For Flory distributed Gaussian chains the form factor becomes \cite{BJSOBW07} 
$F(k) = 2/((a k)^2+\mu)$
in the intermediate wavevector regime. Since the first term in
eq~\ref{eq_veffgen} can again be neglected in the incompressible limit
($\xi k \ll 1$), this yields an effective excluded volume
\begin{equation} 
\veff(k) \approx \frac{1}{F(k)\rho} = \frac{1}{2\rho} ((a k)^2 + \mu), 
\label{eq_veffEP}
\end{equation}
i.e. the term $(a  k)^2$ in eq~\ref{eq_veffasym} is replaced by $(a k)^2+\mu$.
Since this applies also for the propagator,
which becomes $G(k) = 1/((a k)^2+\mu)$,
the central eq~\ref{eq_veff_g} remains valid for Flory distributed chains and, hence, 
also the sum rule eq~\ref{eq_Pr_sumrule}.
We compute as before eq~\ref{eq_Ione_b} using now
the inverse Fourier transform of $G(k)$ in real space
\begin{equation}
G(r) = \frac{1}{4\pi a^2 r} e^{-\sqrt{\mu} r/a}.
\label{eq_gr_EP}
\end{equation}
This yields 
$\Pone(r) = \Ione(r)/G(r) = \Poneasym(r) \ \hone(x)$
with $\Poneasym(r)$ denoting the asymptotic power law eq~\ref{eq_P1rkey}
(with $\cinf=1$) and
$$\hone(x) = \left(1 + 2x \right)^2 \exp(- 2x)$$
a rapidly decaying function 
of $x \equiv \sqrt{\mu} r/ 2a  \approx r/R$
with $R \approx b \Nav^{1/2}$ being the typical end-to-end distance of the polydisperse system.

Interestingly, the diagram (1) is not sufficient to characterize the bond-bond
correlation for larger distances since the last diagram (2) of
Figure~\ref{fig_graph}(b) corresponding to the convolution integral
\begin{eqnarray}
\Itwo(\qvec) & = &
    \int \frac{d\kvec}{(2\pi)^3} G(\qvec) \Acal(\qvec,-\kvec)) G(\qvec-\kvec) 
   \nonumber \\
   & & \hspace*{1.3cm} \times [-\veff(\kvec)] G(\kvec).
   \label{eq_Itwo} 
\end{eqnarray}
provides, as we shall see, the actual cut-off of the power law in this limit. 
Using again eq~\ref{eq_Acal_F} and eq~\ref{eq_veff_g} 
the integral factorizes 
\begin{eqnarray}
\Itwo(\qvec) 
  & = & \frac{-1}{2\rho} \int \frac{d\kvec}{(2\pi)^3} G(\kvec) 
       \times \underline{G(\qvec) \frac{l^2 \qvec^2}{9}}
       \label{eq_Itwo_a}
\\
 & \equiv & - \ctwo \times \ (a \qvec)^2 G(\qvec)
   \label{eq_Itwo_b}
\end{eqnarray}
where we have introduced in the last line the convenient dimensionless constant
\begin{equation}
\ctwo = \frac{(l/a)^2}{18\rho a^3} \int \frac{d\kvec}{(2\pi)^3} a^3 G(\kvec)
\label{eq_ctwo_def}
\end{equation}
in which we dump local physics at large wavevector $\kvec$.
%
%
%
Before evaluating the angular correlations in real space
it is important to clarify the physics described by the diagram. 
The underlined second factor in eq~\ref{eq_Itwo_a} 
characterizes the alignment of the bond vectors of the monomers $n_0$ and $n_1-1$ at a fixed
distance $\rvec$ of the monomers $n_0$ and $n_1=n_0+s$ 
(Figure~\ref{fig_graph}(a)). Obviously, even for Gaussian chains
these two bonds become more and more aligned if the distance $r=|\rvec|$ 
gets larger than $b s^{1/2}$, i.e. when the chain segment becomes stretched.
For perfectly Gaussian chains the bonds $\lnulvec$ and $\lonevec$ at $n_0$ and $n_1$ would
still remain uncorrelated, however, since the second bond is outside 
the chain segment on which we have imposed the distance constraint.

As indicated by the dashed line in the diagram, it is then due to the effective
interaction between the monomers within the stretched segment ($n < n_1$) 
and the monomers outside ($n > n_1$) that the bonds at $n_1-1$ and $n_1$
get aligned and then in turn the two bonds at $n_0$ and $n_1$.
%
%
We note that, strictly speaking, $\ctwo$ depends on the mean chain
length $\Nav$, since $G(k)$ is a function of $\mu$. However, one checks readily
that this effect can be neglected for reasonable mean chain lengths.
%
We also note that the constant $\ctwo$ is {\em finite}, since the UV divergence 
which formally arises for large $k$ (where $\ctwo \sim k$) may be regularized 
by local and, hence, model dependent physics \cite{foot_cutoff}. 
We will determine $\ctwo$ numerically from our simulations of self-assembled
linear equilibrium polymers (Section~\ref{sec:simu}).

%
Assuming a finite and chain length independent coefficient $\ctwo$ in eq~\ref{eq_Itwo_b}
and inserting the propagator $G(q) = 1/((a q)^2+\mu)$ for Flory distributed chains
we obtain by inverse Fourier transformation
\begin{equation}
\Itwo(r) = \ctwo \ (\mu G(r) - \delta(r))
\label{eq_Itwo_c}
\end{equation}
for the interaction integral in real space.
Normalizing $\Itwo(r)$ as before with $G(r)$ and summing over both diagrams 
this yields 
\begin{equation}
\Pone(r)  =  \Poneasym(r) \hone(x) + \ctwo \mu
\label{eq_EPcorr}
\end{equation}
for $r \gg \xi > 0$.
Comparing both terms in eq~\ref{eq_EPcorr} one verifies that the crossover
occurs at $\rstar \approx b \Nav^{1/3}$ in agreement with eq~\ref{eq_rstar}.
The bond-bond correlation function of an incompressible solution of 
Flory distributed polymers becomes thus constant for $r \gg \rstar$.
This remarkable result is essentially due to the polydispersity.
This allows to find for all distances $r$ pairs of bonds $\lnulvec$ and $\lonevec$ 
stemming from segments which are slightly stretched by an energy of order $\mu \ll 1$ and
which are, hence, slightly shorter than a unstretched segment of length $s \approx (r/b)^2$.
Since there are more shorter chains and chain segments this just compensates the
decay of the weight due to the weak stretching.
Although the number of such slightly stretched segments decays strongly with distance, 
their {\em relative} effect with respect to the typical unstretched segments,
$e^{\mu}-1 \approx \mu$, remains constant for all $r$.
It is for this reason that the chemical potential appears in the second term of eq~\ref{eq_EPcorr}. 
Please note that bond pairs from strongly 
stretched segments (corresponding to an energy much larger than $\mu$) are, however, still 
exponentially suppressed and can be neglected. 
As we will show now, this is different for monodisperse chains where strongly 
stretched chain segments contribute increasingly to the average for large distances.
%

\paragraph{Finite chain size effects: Monodisperse chains.}
The bond-bond correlation function of monodisperse polymer melts 
may be obtained by inverse Laplace transformation of the result obtained for Flory 
distributed grand-canonical polymers. Note that a very similar calculation has been 
described in detail in ref~\citen{BJSOBW07} for the coherent structure factor.
The interaction integral $I_N(\qvec)$ for monodisperse chains of chain length $N$
in reciprocal space may we written
\begin{equation}
I_{N}(\qvec) = \Lcal^{-1}\left[ \mu^{-2} \left( \Ione(\qvec,\mu) + \Itwo(\qvec,\mu) \right)\right]
\label{eq_poly2mono}
\end{equation}
where $f(\mu)=\Lcal[f(N)] = \int_0^{\infty} f(N) e^{-N \mu}$ denotes the Laplace transform
of a function $f(N)$. 
The extra factor $\mu^{-2}$ stands for the dangling tails in both diagrams
which accounts for the combinatorics necessary due to the finite chain length.
$\Ione(\qvec,\mu)$ and $\Itwo(\qvec,\mu)$ are the interaction integrals computed 
in the previous paragraph using eq~\ref{eq_veffEP} for the effective interaction 
potential $\veff(k,\mu)$ and the corresponding Fourier-Laplace transformed propagator 
$G(k,\mu)=1/((a k)^2+\mu)$.
Note that the first term in eq~\ref{eq_poly2mono} is accurate up to finite-size corrections 
due to the use of eq~\ref{eq_veffEP} for the effective potential.

We compute then the inverse Fourier transformation $I_N(r) = \Fcal^{-1}\left[I_N(\qvec)\right]$ 
and normalize it consistently with 
$\Fcal^{-1}\left[ \Lcal^{-1}\left[\mu^{-2} G(q,\mu) \right] \right]$ 
using eq~\ref{eq_gr_EP}. This yields 
\begin{equation}
\Pone(r) = \Poneasym(r) \  \hone(x) + \frac{\ctwo}{N} \ \htwo(x)
\label{eq_Ncorr}
\end{equation}
with $x \equiv r/(2a \sqrt{N})$ and $\ctwo$ defined as for Flory distributed chains
(eq~\ref{eq_ctwo_def}).
The first term due to diagram (1) of Figure~\ref{fig_graph}(b) contains a 
(rapidly decaying) cut-off function 
$$\hone(x) = \frac{\mbox{i}^2\mbox{erfc}(2 x)+ 2 x \ \mbox{i}^1\mbox{erfc}(2 x) + x^2 \mbox{erfc}(2x)}{\mbox{i}^2\mbox{erfc}(x)}$$
with $\mbox{i}^n\mbox{erfc}(x)$ denoting the repeated integral of the 
complementary error function \cite{abramowitz}.
This function is non-monotic and goes through a maximum with an overshoot of about 
$54\%$ at $x \approx 0.39$.
The second term in eq~\ref{eq_Ncorr} is again due to diagram (2).
The function
$$\htwo(x) = \frac{\mbox{erfc(x)}}{4 \mbox{i}^2 \mbox{erfc}(x)}$$
becomes constant for small $x$ where $\htwo(x) \approx 1+ 2x/\sqrt{\pi}$.
We note that as for Flory distributed chains the second term in eq~\ref{eq_Ncorr}
becomes dominant on scales $r \gg \rstar(N) \approx b N^{1/3}$.
Interestingly, for large $x$ we have $\htwo(x) \sim x^2 -1/2$ 
and $\Pone(r)$ is, hence, predicted to {\em increase} as
\begin{equation}
\Pone(r) \approx \frac{\ctwo}{N} x^2 \sim (r/N)^2 \mbox{ for } x \gg 1.
\label{eq_rlarge_mono}
\end{equation}
We remind that for a chain segment of arc-length $s$ stretched between its
end monomers $n_0$ and $n_1$ one expects
$\la \lnulvec \cdot \lonevec \ra \approx \ctwo (r/s)^2$
if $r \gg b s^{1/2}$. The coefficient $\ctwo$ stands for the 
correlation of the bond at monomer $n_1-1$ within the stretched segment
with the bond $\lonevec$ at the monomer $n_1$ outside the segment (Figure~\ref{fig_graph}(b)).
The limiting behavior, eq~\ref{eq_rlarge_mono}, is thus expected since
more and more bond pairs from strongly stretched chain segments with $s \to N$ 
must contribute to the average at distances $r \gg b N^{1/2}$. 
%

\paragraph{Finite persistence length effects.}
Up to now we have supposed that the chains are perfectly flexible.
For semiflexible chains ($\cinf > 1$) the above results apply now to 
the Kuhn segments of the chains \cite{RubinsteinBook}.
%
%
The bond length $l$ of the Gaussian reference chain used in the perturbation calculation 
corresponds to the length of the Kuhn segment $\lKuhn = b \sqrt{\cinf} = l \cinf$
(i.e. {\em not} to the effective bond length $b$),
the arc-length $s$ to the number of Kuhn segments $\sKuhn = s/\cinf$ and
the density $\rho$ to the density of Kuhn segments 
\begin{equation}
\rho \Rightarrow \rhoKuhn = \rho/\cinf.
\label{eq_rhoK}
\end{equation}
If the bond-bond correlation function $\Pone(r)$ calculated in
terms of Kuhn segment units is reexpressed in the natural microscopic units,
eq~\ref{eq_rhoK} introduces the additional prefactor $\cinf$ indicated in eq~(\ref{eq_P1rkey}).

\section{\normalfont \large \bfseries Algorithmic and technical issues}
\label{sec:algo}
\begin{figure}[t]
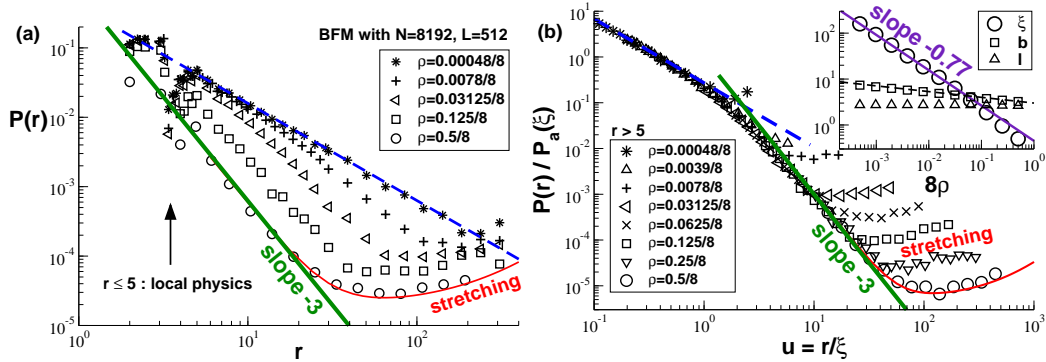

\includegraphics*[width=0.9\colsize]{fig_rhosimple}
\includegraphics*[width=0.9\colsize]{fig_rhoscal}
\caption{Bond-bond correlation function $\Pone(r)$ obtained 
for one chain length 
$N=8192$ and various monomer densities $\rho$ as indicated in the panels:
{\bf (a)} Unscaled raw data,
{\bf (b)} rescaled correlation function 
$\Pone(r) / \Poneasym(\xi)$
versus the natural scaling variable $u=r/\xi$ using the
length scales $\xi(\rho)$, $\be(\rho)$ and $l(\rho)$ indicated in the inset
of the panel.
%
The predicted power-law asymptotes for dense (eq~\ref{eq_P1rkey}) and dilute
(eq~\ref{eq_P1rdilute}) systems are indicated by bold and dashed lines,
respectively. The increase of $\Pone(r)$ visible for large distances 
due to the enhanced weight of stretched chain segments 
is well described by eq~\ref{eq_Ncorr} 
as indicated for $\rho=0.5/8$ (thin lines) assuming $\ctwo=0.14$.
}
\label{fig_rho}
\end{figure}

%
The theoretical predictions derived above are supposed to hold in any 
dense polymer solution containing sufficiently long chains.
The numerical data presented below (Figures~\ref{fig_rho}-\ref{fig_EP})
have been obtained using the well-known ``bond fluctuation model" (BFM) 
\cite{BFM,Deutsch,Paul91a,WBM07,papRouse,WCK09,WMC98b} 
--- an efficient lattice Monte Carlo scheme where a coarse-grained monomer 
occupies 8 lattice sites on a simple cubic lattice 
(i.e., the volume fraction is $8\rho$) and bonds between adjacent monomers 
can vary in length and direction \cite{foot_beadspring}.
All length scales are given in units of the lattice constant and we set $\kB = T =1$.
Even the partial overlap of monomers is forbidden in the classical formulation
of the BFM \cite{BFM,Deutsch,WBM07,WMC98b}. 
We have put the predictions to a test by simulating systems having either 
a quenched and monodisperse or an annealed size distribution: 

{\em (i)}
Monodisperse systems have been equilibrated 
and sampled using a mix of local, slithering snake, and double bridging 
Monte Carlo moves.
See ref~\citen{WBM07} for details.
Systems with chain lengths up to $N=8192$ have been obtained for various densities,
as indicated in Figure~\ref{fig_rho}, up to the ``melt density" $\rho=0.5/8$ \cite{WBM07}.
We use periodic simulation boxes of linear size $L=512$.
For the largest density used this corresponds to
$\nmon = 2^{23} \approx 10^7$ monomers 
and to $1024$ chains of length $N=8192$ per simulation box. 
The smallest density indicated ($\rho=0.00048/8$) refers to one single chain in the box
allowing us to characterize properly the dilute reference point.
The scaling of bond-bond correlation function with chain length $N$ at $\rho=0.5/8$
is presented in Figure~\ref{fig_N}.

{\em (ii)}
As sketched in Figure~\ref{fig_EP}, systems with annealed size distribution
--- so-called ``equilibrium polymers" --- have been obtained 
by attributing a finite and constant scission energy $E$ to each bond
which has to be paid whenever the bond between two monomers is broken.
Standard Metropolis Monte Carlo is used to reversibly break and recombine the chains
\cite{WMC98b,BJSOBW07,WBCHCKM07}. 
Branching and formation of closed rings are explicitly forbidden.
As one expects from standard linear aggregation theory, 
the density of chains $\rho_N$ shows essentially a  Flory distribution,
$\rho_N \sim \exp(-N/ \Nav)$,
with the mean chain size $\Nav$ scaling as
$\Nav^2 \sim \rho \exp(E/\kB T)$ \cite{WMC98b}.
Only local hopping moves have been used for sampling equilibrium polymer systems,
since the breaking and recombination of chains reduce the relaxation times dramatically 
compared to monodisperse systems \cite{HXCWR06}.

\section{\normalfont \large \bfseries Simulation results}
\label{sec:simu}

\begin{figure}[t]
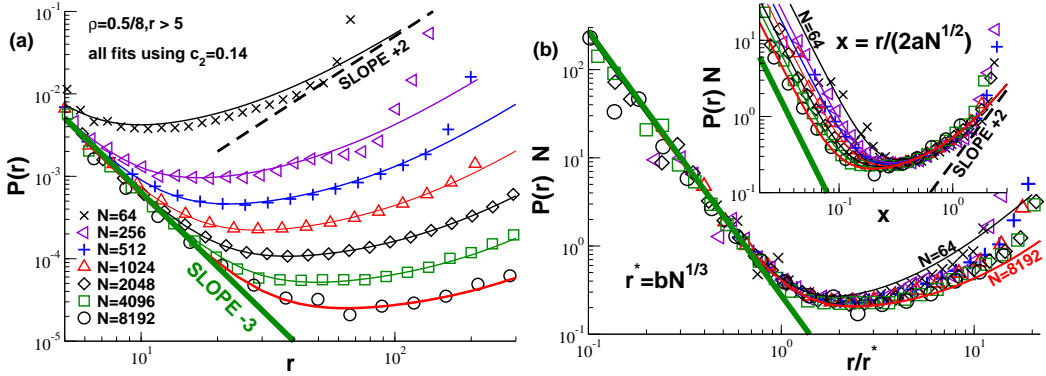

\begin{center}
\includegraphics*[width=0.9\colsize]{fig_Nsimple}
\includegraphics*[width=0.9\colsize]{fig_Nscal}
\caption{$\Pone(r)$ for $r > 5$ for different chain lengths $N$ 
at melt density $\rho=0.5/8$.
The thin lines indicate eq~\ref{eq_Ncorr} for each $N$ assuming $\ctwo=0.14$.
The limiting behavior for large distances, eq~\ref{eq_rlarge_mono}, 
is given by the dashed lines.
{\bf (a)} The data approach systematically the power-law decay (bold line)
predicted by eq~\ref{eq_P1rkey} with increasing chain length $N$.
{\bf (b)} $\Pone(r) N$ as a function of $r/\rstar$ assuming eq~\ref{eq_rstar}
(main panel) and as a function of $x = r/(2aN^{1/2})$ (inset).
The first scaling variable scales the data around the minimum of $\Pone(r)$,
the second for larger distances.
}
\label{fig_N}
\end{center}
\end{figure}

\begin{figure}[t]
\begin{center}
\includegraphics*[width=0.9\colsize]{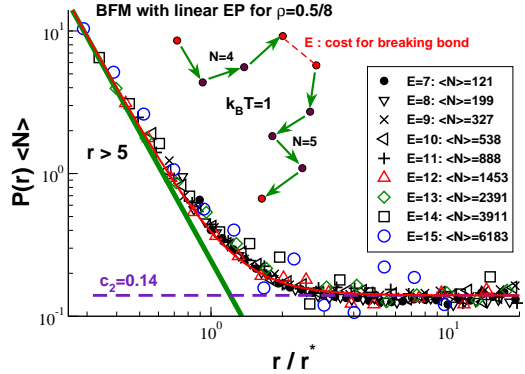}
\caption{Bond-bond correlation function $\Pone(r)$ for equilibrium polymers at density
$\rho=0.5/8$ of the BFM for different bond energies $E$.
This energy has to be paid for the (reversible) breaking of a bond as indicated by the 
dashed line in the sketch where a chain of $N=9$ monomers is broken into two chains of 
length $N=4$ and $N=5$.
The corresponding mean chain lengths $\la N(E) \ra$ are indicated.
If $\Pone(r) \Nav$ is traced as a function of $r/\rstar$ with
$\rstar = b \Nav^{1/3}$ all data points collapse.
Note that $\Pone(r) \Nav \to \ctwo= 0.14$ for $r/\rstar \gg 1$.
The complete prediction eq~\ref{eq_EPcorr} (thin line) interpolates perfectly between the
power-law asymptote (bold line) and the plateau (dashed line).
}
\label{fig_EP}
\end{center}
\end{figure}

\paragraph{Density effects.}
The scaling of the bond-bond correlation function with density for monodisperse chains 
is addressed in Figure~\ref{fig_rho}. 
Only data for our largest chain length $N=8192$ is presented here to focus first on the 
discussion of the large-$N$ limit. The finite-$N$ effects, visible nevertheless (``stretching") 
from the increase of $\Pone(r)$ for large $r$, will be further discussed below at the end of 
this section (Figures~\ref{fig_N} and \ref{fig_EP}).
As can be seen in panel (a), model-depending physics not taken into account by the theory 
obviously becomes relevant for short distances corresponding to segments of a couple of monomers. 
For clarity we have thus omitted data points with $r \le 5$ in all other figures and panels below.

The power-law behavior observed for small densities is implicit to the swollen-chain statistics
in dilute good solvents where the root-mean-square segment size $R(s) = \bdil s^{\nudil}$
is known to scale with a Flory exponent $\nudil \approx 0.588$ 
(with $\bdil$ denoting the respective power-law amplitude).
Since $\Pone (s) \sim \partial^2_s R^2(s) \sim s^{-\omegadil}$ with
$\omegadil = 2-2\nudil$ (see ref~\citen{WMBJOMMS04}) 
it follows from eq~\ref{eq_sr_scaling} that
the bond-bond correlation function should scale as 
\begin{equation}
\Pone(r) \approx \Ponedil(r) \equiv (\bdil/l)^2 (r/\bdil)^{-\omegadil/\nudil} \sim r^{-1.40}
\label{eq_P1rdilute}
\end{equation}
as indicated by the dashed lines in Figure~\ref{fig_rho}.
The index $0$ indicates that this is the expected asymptotic behavior for asymptotically
long chains in the dilute good solvent limit.

As the density increases, the correlation function still coincides with the dilute behavior 
for distances $r$ smaller than the screening length $\xi(\rho)$ of the semidilute solution 
where each chain segment interacts primarily with itself. 
(The screening length is indicated in the inset of panel (b).)
At distances $r \gg \xi(\rho)$, where the chains overlap and form a ``melt of blobs", 
\cite{DegennesBook} the correlation function decreases much faster, however not exponentially 
as one might expect from Flory's hypothesis or the $\delta(r)$-scenario (``self-kicks") 
mentioned in the Introduction, but with a second power-law regime with exponent $\omega/\nu=3$.

That the crossover between both power-law regimes occurs indeed at $r \approx \xi(\rho)$
can better be seen from panel (b) where we have replotted the data as a function of the 
reduced distance $r/\xi(\rho)$ as suggested by eq~\ref{eq_P1scaling}. Taking apart the
finite chain-size effects for large distances we find an excellent scaling collapse of
the data considering the large spread of the raw data (panel (a)) and that no free shift 
parameter was used.
The three independently determined length scales $\xi$, $b$ and $l$ used for the rescaling
according to eq~\ref{eq_P1scaling} are compared in the inset of panel (b).
The latter two lengths are relevant due to the stiffness parameter $\cinf = (b(\rho)/l)^2$ 
which is needed (eq~\ref{eq_P1rkey}) for the vertical scale $\Poneasym(\xi)$.
Note that the bond length $l$ (triangles) is essentially constant for all densities,
at least on the logarithmic scales addressed here. The effective bond length $b$
approaches $l$ from above, i.e. $\cinf(\rho)$ decreases with increasing $\rho$.
The dashed line in the inset corresponds to the power law expected 
from the scaling theory of semidilute solutions \cite{DegennesBook}
\begin{equation}
b^2(\rho) \sim \rho^{-(2\nudil-1)/(3\nudil-1)} \sim \rho^{-0.23}.
\label{eq_berho}
\end{equation}
Thus, the density dependence of $\cinf$ can not be neglected.
The screening length $\xi$ has been determined assuming eq~\ref{eq_vbare}
and using the directly measured effective bond length $b(\rho)$ and 
dimensionless compressibility $g(\rho)$. We have crosschecked these values
with the decay of the total structure factor where this has been possible.
For not too high densities our data is nicely fitted by the power law (bold line)
\begin{equation}
\xi(\rho) \sim  \rho^{-\nudil/(3\nudil-1)} \sim \rho^{-0.77}
\label{eq_xirho}
\end{equation}
as expected for semidilute solutions.\cite{DegennesBook} 
Obviously, the semidilute power-law relations cannot hold strictly for the highest densities
where the compressibility (and, hence, the blobs size) becomes too small
\cite{Paul91a,KL08b}. However, the
differences are small on logarithmic scales and we obtain a very similar scaling plot 
by insisting on eqs~\ref{eq_berho} and \ref{eq_xirho} for all densities (not shown).

The density crossover scaling implies obviously the matching of the dilute and dense 
asymptotic power-law predictions
\begin{equation}
\Ponedil(\xi) \approx \Poneasym(\xi).
\label{eq_PaPdil}
\end{equation}
As can be checked using the well-known scaling relations
$b^2 \approx \xi^2/g$, $\xi^3\rho \approx g$ and $\xi \approx \bdil g^{\nudil}$
defining the semidilute solution \cite{DegennesBook}
(which are also implicit to eq~\ref{eq_berho} and eq~\ref{eq_xirho}),
eq~\ref{eq_PaPdil} requires the prefactor $\cinf(\rho)$ in eq~\ref{eq_P1rkey}.
We have checked that $\Ponedil(\xi) \sim \xi^{-\omegadil/\nudil}$ scales the data,
while $\Poneasym(\xi)$ without $\cinf$ does not (not shown). The successful scaling 
collapse confirms, hence, the rescaling of 
the Kuhn segments presented at the end of section~\ref{sec:theo}.

\paragraph{Finite $N$-effects for monodisperse chains.}
We concentrate in the reminder of this section on dense melts with $\rho=0.5/8$.
Chain length effects are investigated in Figure~\ref{fig_N} for monodisperse chains.
The unscaled raw data are presented in panel (a).
As expected from theory,  $\Pone(r)$ increases for large distances due to the 
increasing weight of stretched segments contributing to the average at distance $r$.
With increasing chain length $N$ this effect becomes less important, however,
and our data approaches systematically the asymptotic power-law decay (eq~\ref{eq_P1rkey})
indicated by the bold line. The thin lines represent the complete 
prediction, eq~\ref{eq_Ncorr}, for different $N$.
Numerical data and theory agree nicely, especially for large chains.
The deviations observed for $N \le 256$ and for fully stretched chains ($r \approx l N$) 
are, of course, expected. 
Please note that the cut-off function $\hone(x)$ of the asymptotic power law 
obtained by diagram (1) may be ignored ($\hone(x)\to 1$) without changing the plot.
The finite-$N$ effects are in fact dominated by the function $\htwo(x)$ 
of the second term in eq~\ref{eq_Ncorr}. 
We remind that eq~\ref{eq_rlarge_mono} predicts ultimately a
quadratic increase with distance of the bond-bond correlation function 
as indicated by the dashed line for one chain length ($N=64$).
It is essentially the limited simulation box size ($L = 512$) of the present study 
preventing us from demonstrating numerically this asymptotic behavior 
which should be accessible otherwise for the longer chains ($N \gg 10^3$). 
%

%
The scaling with chain length is addressed in panel (b).
Obviously, eq~\ref{eq_Ncorr} is not compatible with {\em one} scaling variable
allowing to collapse all data.
The crossover from the power-law asymptote (diagram (1)) to the $N$-dependent correction
(due to diagram (2)) is, however, well described by plotting $\Pone(r) N$ 
as a function of $r/\rstar$ with $\rstar \equiv b N^{1/3}$ 
using as a scale the minimum of eq~\ref{eq_Ncorr}. As shown in the main panel
this yields a numerically satisfactory scaling over two orders in magnitude 
of the reduced distance $r/\rstar$, especially for our largest chains. 
This scaling fails for large distances ($r \gg R(N)$), however, as emphasized 
by the theoretical predictions (thin lines) indicated for $N=64$ and $N=8192$ (bottom).
In this limit only the second term in eq~\ref{eq_Ncorr} matters and,
as can be seen from the inset in Figure~\ref{fig_N}(b), a data collapse 
can be achieved by tracing 
$\Pone(r) N$ as a function of $x \approx r/R(N)$.
As already noted, the observed deviations from the limiting behavior for $x \gg 1$
(eq~\ref{eq_rlarge_mono}) are expected (i) due to the breakdown of the Gaussian chain model 
at $r \approx N l$ for small chain lengths and (ii) due to the restricted box size
of the present study.

\paragraph{Equilibrium polymers.}
Fortunately, this scaling issue is much simpler for equilibrium polymers as shown 
in Figure~\ref{fig_EP} where we have plotted $\Pone(r) \Nav$ 
as a function of $r/\rstar$ with $\rstar \equiv b \Nav^{1/3}$ 
using the indicated mean chain lengths $\Nav$.
Note that the error bars (not shown) become clearly much larger than the 
symbol size for large bond energies $E > 12$. It is fair to state, however,
that all data points collapse nicely on the {\em one} master curve indicated 
by the thin line obtained from eq~\ref{eq_EPcorr} for $\Nav \to \infty$. 
That the variable $x$ effectively drops out stems from 
(i) the rapid decay of the exponential cut-off function of the first term 
in eq~\ref{eq_EPcorr} and (ii) the $x$-independence of the correction term.
That the bond-bond correlation function of equilibrium polymers becomes constant 
for large distances confirms a non-trivial prediction of the theory. 
We used the clearly visible plateau 
to determine the coefficient $\ctwo=0.14$ required for the interaction diagram (2)
and already used in the previous Figures~\ref{fig_rho} and \ref{fig_N}.
We note finally that this best fit value of $\ctwo$ is rather close to
$\Pone(s=1)= \la \lvec_n \cdot \lvec_{n+1} \ra/l^2 \approx 0.10$,
the independently determined bond-bond correlation between adjacent bond vectors \cite{foot_cutoff}.

\section{\normalfont \large \bfseries Conclusion}
\label{sec:summary}
%
\paragraph{Summary.}
Focusing on dense solutions of linear and (essentially) flexible polymer chains
we have investigated analytically and by means of Monte Carlo simulation 
the intrachain angular correlations with respect to the distance $r$ between bond pairs.
Motivated by recent work showing the power-lay decay of the bond-bond correlation function 
$\Pone(s)$ with curvilinear distance $s$ (eq~\ref{eq_Psomega}) we addressed the question
whether the correlations are indeed long-ranged in space or only due to the return probability
(``self-kicks") of a chain.
Our calculations of $\Pone(r)$ have been based on a standard one-loop perturbation scheme 
(Figure~\ref{fig_graph}(b)). The power-law decay predicted for asymptotically long chains 
(eq~\ref{eq_P1rkey}) is well confirmed by our numerically data approaching systematically
the asymptotic envelope $\Poneasym(r)$ with increasing density (Figure~\ref{fig_rho})
and (mean) chain length (Figures~\ref{fig_N} and \ref{fig_EP}).
As postulated in eq~\ref{eq_P1scaling}, density effects are 
found to scale as $\Pone(r)/\Poneasym(\xi) = f(u=r/\xi)$ 
with $\xi(\rho)$ being the independently measured screening length.
For $u \ll 1$ we confirm the expected power law for dilute good solvents (eq~\ref{eq_P1rdilute}). 
More importantly, the explicit compressibility dependence of the bond-bond correlation function
drops out for $u \gg 1$ where $f(u) \Rightarrow 1/u^3$,
i.e. polymer solutions behave on large scales as incompressible packings of blobs
and this for all densities provides that the chains are sufficiently long. 
Finite-chain size effects are also successfully predicted by the theory for both
monodisperse polymers (eq~\ref{eq_Ncorr}) and equilibrium polymers (eq~\ref{eq_EPcorr}).
It should be emphasized that the fit to the asymptotic power law is parameter free
and that fitting the $N$-effect only requires one additional free parameter, 
$\ctwo$ (eq~\ref{eq_Itwo_b}), 
due to local orientational correlations between adjacent bonds
which regularize diagram $(2)$ in Figure~\ref{fig_graph}(b).

\paragraph{Outlook.}
Interestingly, the presented perturbation calculation for dense polymer chains may also 
be of relevance to the angular correlations of dilute polymer chains at and around
the $\Theta$-point which has received attention recently \cite{Rubinstein08}. 
The reason for this connection is that (taken apart different prefactors) the {\em same} 
effective interaction potential (eq~\ref{eq_veffasym}) enters the perturbation
calculation in the low wavevector limit. We expect therefore similar genuinely 
long-ranged angular correlations for asymptotically long $\Theta$-chains 
(as implied by eq~\ref{eq_sr_scaling}) 
rather than the ``self-kicks" suggested in the literature \cite{Rubinstein08}. 
Strong finite-$N$ effects are again expected, however, and much longer chains 
as the ones presented in the current numerical studies of $\Theta$-chains are 
required to demonstrate the asymptotic power-law behavior.

The present study has focused on the first Legendre polynomial of the intrachain
bond-bond correlations \cite{foot_Ponedef}. Since this is not an easily experimentally 
accessible property it should be mentioned that higher Legendre polynomials have also 
been predicted following similar perturbation calculations as the ones presented above. 
For instance it is possible to show that the second Legendre polynomial should decay as the 
fifth power of distance if averaged over all intra-chain contributions and as the sixth power 
if averaged over all bond pairs at a given distance \cite{foot_Pld}.
Unfortunately, we are at present unable to demonstrate these predictions {\em numerically}
due to the stronger power-law decay requiring much better statistics.
This study is currently underway.

Our one-loop perturbation calculation show that for infinite chains or Flory distributed 
(grand-canonical) polymers the bond-bond correlations observed in systems of linear chains 
are equivalent to the subtraction of angular correlations due to closed cycles 
(eq~\ref{eq_Pr_sumrule}). 
%
%
The same sum rule can be shown to hold for higher Legendre polynomials
summing over intra- and inter-chain contributions. This finding suggests a 
deeper purely geometrical description of the observed long-ranged 
orientational correlations relating this issue to the recently discovered 
Anti-Casimir forces in polymer melts \cite{ANS05a,ANS05b} arising due to a 
similar subtraction of soft fluctuation modes, not present in the linear polymer 
system but in its hypothetical counterpart containing both chain architectures.
Taking advantage of the polymer-magnetic analogy\cite{DegennesBook} 
we plan to address this fundamental connection in a more theoretical paper.

\vspace{4mm}

{\bf Acknowledgment.} 
We thank A.N.~Semenov (ICS, Strasbourg, France) for helpful discussions. 
Financial support by the University of Strasbourg, the CNRS, the IUF and the ESF-STIPOMAT programme 
are acknowledged as well as a generous grant of computer time by the IDRIS (Orsay).


\begin{thebibliography}{10}
\bibitem{foot_asympN}
The local properties $l$, $\be$, $g$ or $\xi$ are supposed to be those of
  asymptotically long chains. The determination of their values requires the
  numerically extrapolation using the data obtained from a systematic chain
  length variation. Extrapolation schemes for the dimensionless compressibility
  $g$ and the effective bond length $\be$ are, e.g., discussed in
  ref~\citen{WCK09}.
\bibitem{foot_Ponedef}
Differences between our definition of the bond-bond correlation function and
  the more common first Legendre polynomial $\la \lnulhat \cdot \lonehat \ra$
  using the normalized tangent vector $\lhat_i = \lvec_i/|\lvec_i|$ can be
  shown to be negligible.
\bibitem{foot_bl}
The two lengths $l$ and $b$ (or $a=b/\sqrt{6}$) are nevertheless given if this
  is helpful to remind well-known relations and to indicate where the different
  contributions stem from.
\bibitem{foot_cutoff}
For soft melts with weak bare excluded volume $v$, i.e. $\xi \gg b$, it can be
  shown that an upper cut-off wavevector $\kcut \approx 1/\xi$ regularizes the
  integral over $k$. The coefficient $\ctwo$ becomes \begin{equation} \ctwo =
  \frac{1}{2\pi\rho b^2\xi}. \label{eq_ctwo_soft} \end{equation} Using
  $b=3.244$ and $\xi = 0.5$ for $\rho=0.5/8$ this gives $\ctwo \approx 0.48$.
  This is not that far off the best fit value $\ctwo \approx 0.14$ considering
  that for small $\xi \le b$ one expects $\ctwo$ to be rather determined by the
  model-depending stiffness between adjacent bonds.
\bibitem{foot_beadspring}
We have crosschecked our lattice Monte-Carlo results by molecular dynamics
  simulation of a standard bead-spring model \cite{WMBJOMMS04,WBM07,papRouse}.
  Sampling dense melts with chain lengths up to $N=1024$ we obtain the same
  qualitative behavior albeit the power-law asymptote is less convincingly
  shown as one expects due to the much shorter chain length.
\bibitem{foot_Pld}
Concerning the general Legendre polynomial $P_l(r)$ characterizing the angular
  correlations of bond pairs summing over both intra- and inter-chain
  contributions we find analytically \begin{equation} P_l(r) \approx
  \frac{(-1)^{l+1}}{\rho r^{l d}} \mbox{ for } \xi \ll r \ll \rstar(N) \sim
  N^{1/d} \nonumber \end{equation} with $l$ being the degree of the Legendre
  polynomial and $d$ the spatial dimension. The simulation of polymer melts in
  thin films ($d=2$) may thus allow to demonstrate more readily the suggested
  behavior for higher ($l>1$) Legendre polynomials.
\end{thebibliography}

\clearpage
\newpage
\onecolumn
\begin{flushleft}
Title:
{\bf Distance dependence of angular correlations in dense polymer solutions}
\\
Authors:
J.P. Wittmer, A. Johner, S.P. Obukhov, H. Meyer, A. Cavallo, J. Baschnagel\\
\end{flushleft}

\vspace*{0.5cm}

\centerline{\includegraphics*[width=14.0cm]{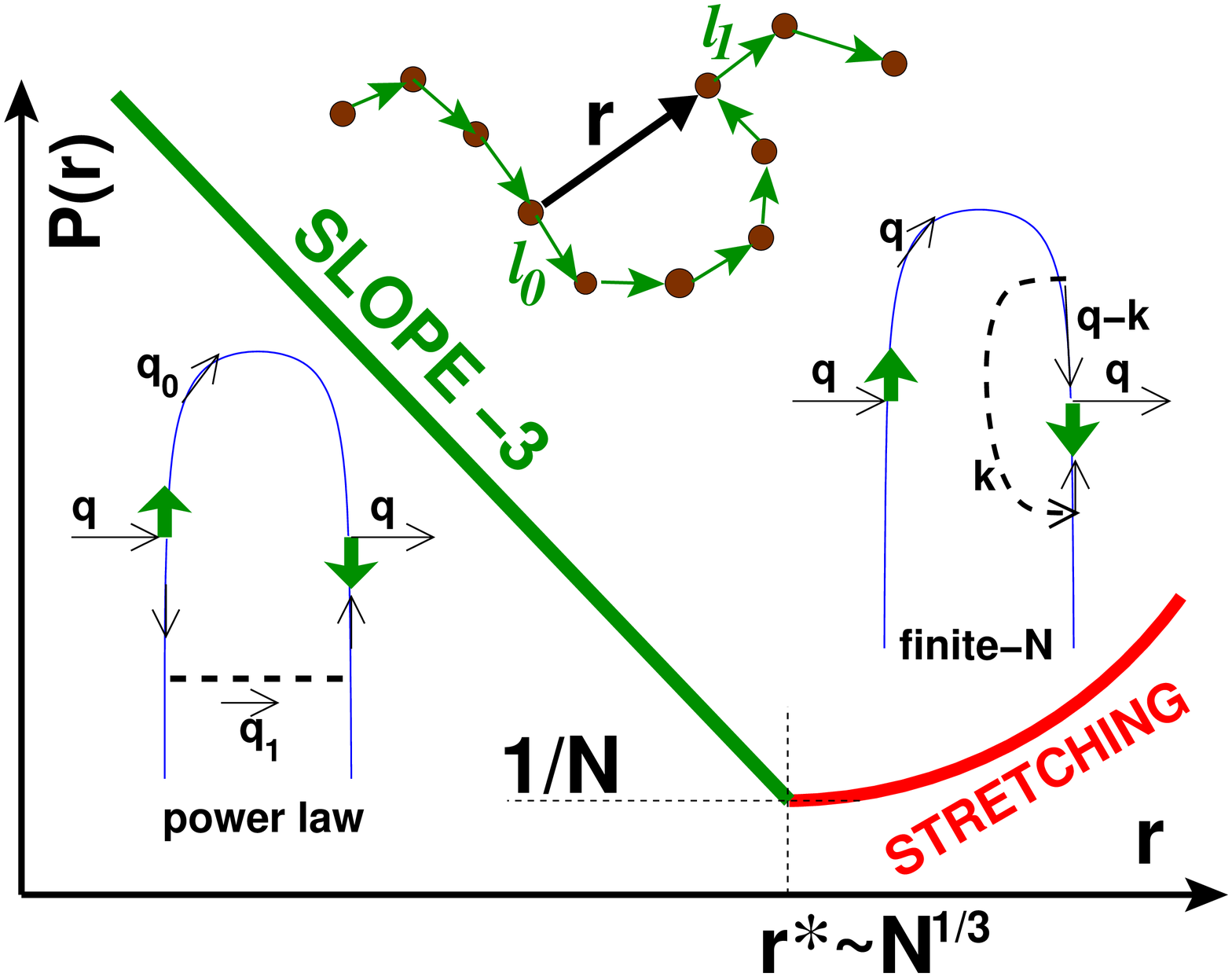}}

\vspace*{0.5cm}
For Table of Contents (TOC) use only.

%

\end{document}